\documentclass[a4paper]{article}
\usepackage{xspace}
\usepackage{amssymb}
\usepackage{amsfonts}
\usepackage{listings}
\usepackage{amsmath}
\usepackage[latin1]{inputenc}
\usepackage[dvips]{graphicx}
\usepackage{pifont}
\usepackage{subfigure}
\usepackage{gastex}
\usepackage{tabls}

\newcommand{\N}{\mathbb{N}}
\newcommand{\IPM}{IPM$(k)$\xspace}

\newcommand{\para}[1]{\left\langle#1\right\rangle}
\newcommand{\og}[1]{\mathcal{G}_{#1}\xspace}
\newcommand{\set}[1]{\left\{#1\right\}}

\newcommand{\ie}{\emph{i.e.}\@\xspace}

\newtheorem{theorem}{Theorem}

\newtheorem{lemma}[theorem]{Lemma}
\newtheorem{proposition}[theorem]{Proposition}

\newtheorem{example}{Example}
\newtheorem{remark}{Remark}
\newcommand{\qed}{}
\newenvironment{proof}{\noindent \emph{Proof.~}}
                      {\rule{0pt}{0pt}\hfill \ding{111}\vspace{2ex}}
\title{On computing fixed points for generalized\\ sandpiles}
\author{E. Formenti\footnote{Laboratoire I3S, Université de Nice-Sophia Antipolis,
2000, route des lucioles, Sophia Antipolis Cedex, \textsc{France}. Email:
\{\texttt{enrico.formenti,benoit.masson}\}\texttt{@i3s.unice.fr}}
 \and B. Masson$^*$}
\date{}
\begin{document}
\lstset{language=C,%
basicstyle=\small,%
showstringspaces=false,%
captionpos=b,%
aboveskip=\bigskipamount%
}
\maketitle

\begin{abstract}
  We prove fixed points results for sandpiles starting with
  arbitrary initial conditions. We give an effective algorithm
  for computing such fixed points, and we refine it in the particular
  case of SPM.
\end{abstract}


\section{Introduction} \label{sec:intro}

Sandpiles are a simple but meaningful formal model for the simulation
of systems governed by self-organized criticality (SOC).  These
systems, starting from any initial configuration, evolve to a
``critical'' state. Any perturbation of this critical state, no matter
how small, originates a deep and uncontrollable reorganization of
the whole.  Then, they start evolving towards another critical state
and so on. SOC systems are commonly used for simulating natural
phenomena like snow avalanches, dune formations, but also woods fires
and even, stock exchange crashes.

The first discrete model for sandpiles, (later on) called SPM
(\emph{Sand Pile Model}), has been introduced in~\cite{BTW}. It is
based on a simple local rule: a sand grain falls on its right if the
difference between its sandpile and the one on its right is bigger
than a certain amount of grains.  In~\cite{GK1,GK2,GMP1,P}, the model
has been mathematically formalized and studied as a discrete dynamical
system.  In particular, they proved that SPM has fixed point dynamics
and exhibited formulas for the precise expression of fixed points and
for computing the transient length.

Similar results exist for a more complete model, \IPM (\emph{Ice Pile
  Model}) introduced in~\cite{GMP1}. The results found for these two
models, synthesized in~\cite{GMP2}, are very interesting and complete
but they concern only very special initial conditions in which all the
sand grains are concentrated in a unique pile and there is no grain
elsewhere.

In this paper we generalize those results to arbitrary initial
conditions. Of course, due to much greater combinatorial complexity of
the problem, we do not have nice formulas but we give a fast algorithm
for computing the fixed point.

\medskip
This paper is structured as follows. The next section recalls basic
definitions about the main sandpile models. Section~\ref{sec:results}
resumes the main known results, which are generalized to arbitrary
sandpiles in Section~\ref{sec:algo}. Section~\ref{sec:spm} improves
the results of the previous section for the special case of SPM. In
the final section we draw our conclusion and give some perspectives.


\section{Basic definitions} \label{sec:def}

A \emph{sandpile} is a finite sequence of integers $(a_1, \ldots,
a_l)$; $l\in\N$ is the \emph{length} of the pile.  Sometimes a
sandpile is also called a \emph{configuration}.  Given a sandpile
$(a_1, \ldots, a_l)$, the integer $n=\sum_{i=1}^l a_i$ is the
\emph{number of grains} of the pile.  Given a configuration $(a_1,
\ldots, a_l)$, a subsequence $a_i,\dots,a_j$ (with $1\leq i,j\leq l$
and $i<j$) is a \emph{plateau} if $a_k=a_{k+1}$ for $i\leq k<j$; a
subsequence $a_i,a_{i+1}$ is a \emph{cliff} if $a_i-a_{i+1}\geq 2$.

In the sequel, each sandpile $(a_1, \ldots, a_l)$ will be conveniently
represented on a two dimensional grid where $a_i$ is the grain content
of column $i$.

\begin{figure}[!ht]
  \centering
  \subfigure[Height differences.]{
    \input{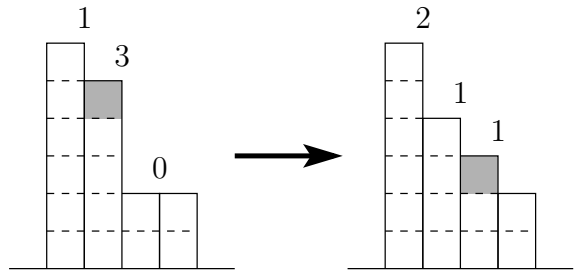}
    \label{fig:fonct_spm_a}
  }\\
  \subfigure[Column grain content.]{
    \input{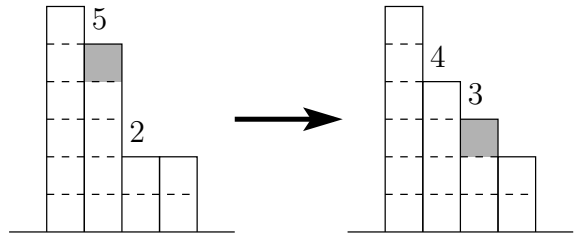}
    \label{fig:fonct_spm_b}
  }
  \caption{updating in the SPM model.}
  \label{fig:fonct_spm}
\end{figure}

\medskip
A \emph{sandpile system} is a finite set of rules that tell how the sandpile
is updated. SPM is the most known and the most simple sandpile system.
It consists in just one local rule. Moreover, all initial configurations
contain $n$ grains in the first column and nothing elsewhere \ie
they are of type $(n)$. The system rule can be defined
in two equivalent ways. The former, introduced in~\cite{BTW},
considers the sequence $z_i = a_i-a_{i+1}$ of differences between two
consecutive columns $i$ and $i+1$. If $z_i \geq 2$, then a sand grain
falls from column $i$ to $i+1$ giving the following new sequence of 
differences (see Figure~\ref{fig:fonct_spm_a}):
$$\left\{\begin{array}{rcl}
z'_{i-1} &=& z_{i-1}+1\\
z'_i &=& z_i-2\\
z'_{i+1} &=& z_{i+1}+1\enspace.
\end{array}\right.$$

The latter, introduced in~\cite{GK2}, deals with the real height of
consecutive co\-lumns. It has the advantage of having a simple and
intuitive graphical representation. The updating rule is defined as
follows (see Figure~\ref{fig:fonct_spm_b})
$$\left\{\begin{array}{rcl}
a'_i &=& a_i-1\\
a'_{i+1} &=& a_{i+1}+1\\
\end{array}\right.\qquad \textrm{if}\;a_i-a_{i+1} \geq 2\enspace.$$

Remark that there are also two ways of evolving the sandpile system:
parallel and sequential execution mode. In parallel mode, all
applicable rules are applied at once; only one rule at a time is
applied in the sequential mode. For example, in
Figure~\ref{fig:fonct_spm}, several grains might fall at the same time
(columns $2$ and $4$ may lose a grain): the next configuration depends
on the execution mode; if we choose the sequential mode, then the next
configuration depends also on the column to which one applies the
system rule.  The execution mode is fixed from the beginning and does
not change along the evolution of the system. In the sequel, we will
be mainly interested in the sequential mode for the simplicity sake.

\smallskip
\IPM is another usual model that extends SPM. It contains the
\emph{vertical} rule of SPM defined above, plus a \emph{horizontal}
rule: the grains can slide on a horizontal plateau of length at most
$k$ as follows
$$
(\ldots, p+1, \underbrace{p, p, \ldots, p}_{k'\;\textrm{times}\;\,
  (k'\leq k)}, p-1, \ldots) \xrightarrow{\quad \textrm{\IPM}_{hor}
  \quad} (\ldots, p, p, p, \ldots, p, p, \ldots)\enspace.$$
This rule can be used to simulate slopes of less than 45°, while the
vertical rule (when generalizing the definition of cliffs to any
difference of height) simulates slopes bigger than 45°.

\medskip
A sequence of configurations $\set{c_i}_{i\in\N}$ is called an
\emph{orbit} of initial condition $c_1$ if for all $i\in\N$, $c_{i+1}$
is obtained from $c_i$ via the application of a system rule.  Remark
that there might be more than one orbit for the same initial
condition.

The set of orbits $O_c$ with the same initial condition $c$ is
graphically represented by the \emph{orbit graph}
$\mathcal{G}_c=\para{V,E}$, where $V$ is the set of all configurations
belonging to orbits in $O_c$ and $(a,b)\in E$ ($E\subseteq V\times V$)
if and only if $b$ is obtained from $a$ by an application of a system
rule.  Denote $\og{c}^l$ the orbit graph of the configuration $c$ in
which all configurations (including $c$) have length at most $l$.
Given an orbit graph $\mathcal{G}=\para{V,E}$ we say that a
configuration $c$ belongs to $\mathcal{G}$, denoted $c\in\mathcal{G}$
if $c\in V$.  A vertex $v\in V$ is a \emph{fixed point} of a directed
graph $G=\para{V,E}$ if $v$ has no outgoing edges.  Given two graphs
$G_1=\para{V_1,E_1}$ and $G_2=\para{V_2,E_2}$, $G_1$ is a
\emph{sub-graph} of $G_2$ if $V_1\subseteq V_2$ and $E_1\subseteq
E_2$.

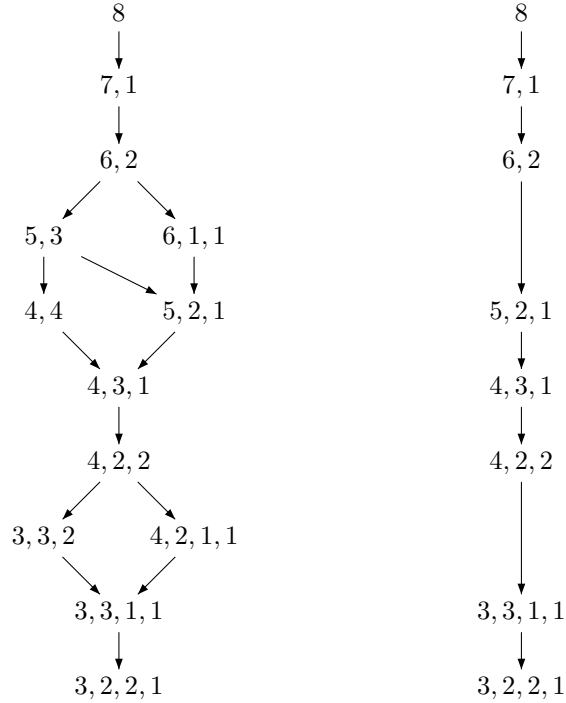
\begin{figure}[!ht]
  \centering
  \subfigure[Sequential execution mode.]{
    \begin{picture}(50,90)(-5,10)
      \gasset{Nframe=n,Nw=10,Nh=5,Nmr=0}

      \node(n0)(20,100){$8$}
      \node(n1)(20,90){$7,1$}
      \drawedge(n0,n1){}
      \node(n2)(20,80){$6,2$}
      \drawedge(n1,n2){}
      \node(n3)(10,70){$5,3$}
      \drawedge(n2,n3){}
      \node(n4)(30,70){$6,1,1$}
      \drawedge(n2,n4){}
      \node(n5)(10,60){$4,4$}
      \drawedge(n3,n5){}
      \node(n6)(30,60){$5,2,1$}
      \drawedge(n3,n6){}
      \drawedge(n4,n6){}
      \node(n7)(20,50){$4,3,1$}
      \drawedge(n5,n7){}
      \drawedge(n6,n7){}
      \node(n8)(20,40){$4,2,2$}
      \drawedge(n7,n8){}
      \node(n9)(10,30){$3,3,2$}
      \drawedge(n8,n9){}
      \node(n10)(30,30){$4,2,1,1$}
      \drawedge(n8,n10){}
      \node(n11)(20,20){$3,3,1,1$}
      \drawedge(n9,n11){}
      \drawedge(n10,n11){}
      \node(n12)(20,10){$3,2,2,1$}
      \drawedge(n11,n12){}
    \end{picture}
    \label{fig:spm8_a}
  }
  \subfigure[Parallel execution mode.]{
    \begin{picture}(50,90)(5,10)
      \gasset{Nframe=n,Nw=10,Nh=5,Nmr=0}

      \node(n0)(30,100){$8$}
      \node(n1)(30,90){$7,1$}
      \drawedge(n0,n1){}
      \node(n2)(30,80){$6,2$}
      \drawedge(n1,n2){}
      \node(n5)(30,60){$5,2,1$}
      \drawedge(n2,n5){}
      \node(n7)(30,50){$4,3,1$}
      \drawedge(n5,n7){}
      \node(n8)(30,40){$4,2,2$}
      \drawedge(n7,n8){}
      \node(n11)(30,20){$3,3,1,1$}
      \drawedge(n8,n11){}
      \node(n12)(30,10){$3,2,2,1$}
      \drawedge(n11,n12){}
    \end{picture}
    \label{fig:spm8_b}
  }
  \caption{orbit of SPM for $n=8$ with respect to execution mode.}
  \label{fig:spm8}
\end{figure}

As an example, Figure~\ref{fig:spm8} shows the orbit for the initial
configuration $(8)$ according to both sequential and parallel
execution modes for SPM. One can remark that both sequential and
parallel execution mode lead to the same fixed point $(3,2,2,1)$. The
difference between the two evolutions seems to consist only in the
transient length. These remarks are true and will be justified in the
following section.

\section{Known results} \label{sec:results}

The results of this section are essentially taken from~\cite{GK2}
and~\cite{GMP1}. We stress that, for simplicity sake, only the  
sequential execution mode is considered. 

\subsection{Fixed points}

First we recall the results for SPM, and for the slightly more complex
model \IPM.

The following theorem shows that, for any initial configuration $(n)$, 
the orbit graph of SPM has a very special structure.

\begin{theorem}[\cite{GK2}]\label{th.spmtrellis}
  For any integer $n$, the orbit graph $\og{(n)}$ for SPM is a lattice
  and is finite.
\end{theorem}

As a consequence of Theorem~\ref{th.spmtrellis} we have that SPM has
fixed point dynamics starting from any configuration $(n)$.  Moreover,
this fixed point is unique and it is reached regardless of the order
in which transitions are made and regardless of the execution mode
(parallel or sequential) that has been chosen. The following lemma
characterizes the elements of the lattice.

\begin{lemma}[\cite{GMP1}] \label{lemm:cond1}
  Consider a configuration $c$ and let $n$ be its number of grains.
  Then, $c\in\og{(n)}$ for SPM if and only if it is decreasing and
  between any two plateaus there is at least a cliff.
\end{lemma}

\begin{remark}
  Consider a configuration $c$ and let $n$ be its number of grains.
  Assume that $c$ contains a plateau of length $3$. Such a plateau can
  be seen as two consecutive plateaus of length $2$. Thus, by
  Lemma~\ref{lemm:cond1}, $c$ does not belong to $\og{(n)}$.
\end{remark}

The previous condition is not necessary for the characterization of
the fixed point. Notwithstanding, it allows to obtain a much simpler
proof of the following Theorem~\ref{th:ipmtreillis}. 

\paragraph{Notation.}
A couple of integers $\para{p,k}$ is the decomposition of $n\in\N$
in its \emph{integer sum} if $n=k+\sum_{i=1}^p i=k+\frac{p\cdot(p+1)}{2}$.

\begin{theorem}[\cite{GK2}] \label{th:ipmtreillis}
  There exists a unique decomposition of $n\in\N$ in its integer sum. 
  Then, the fixed point $\Pi$ obtained starting from the initial
  configuration $(n)$ is the following  
  $$\Pi = \left\{\begin{array}{l@{\hspace{1cm}}l}
      (p, p-1, \ldots, 1) & \textrm{if}\; k=0\\
      (p, p-1, \ldots, k+1, k, k, k-1, \ldots, 1) & \textrm{otherwise}
      \enspace .
    \end{array}\right.
  $$
\end{theorem}

\medskip
Similar results hold for \IPM, we recall the main ones here. They can
be found in their original form in~\cite{GMP1}.

\begin{theorem}[\cite{GMP1}]\label{th.ipmtrellis}
  For any integer $n$, the orbit graph $\og{(n)}$ for \IPM is a
  lattice and is finite.
\end{theorem}

Again, there exist a characterization of the configurations of the
lattice \cite{GMP1}. It is a generalization of Lemma~\ref{lemm:cond1}
which will not be used explicitly here. Remark that this allows to
give the exact form of the fixed point of a configuration for \IPM.

\subsection{Transient length}
We have seen that according to the usual models, the sandpile $(n)$
evolves to a fixed point. It can be interesting to know how much
time (\ie how many iterations of the system rule) it takes to reach
such a fixed point - of course this time depends on the execution mode.

For SPM, in the sequential execution mode, it is easy to compute the
number of steps needed to reach the fixed point: it suffices to remark
that the grains in the $i$-th column took $i$ iterations to reach it;
taking the sum all over the number of grains of the fixed point one
finds the number of iterations.

\begin{theorem}[\cite{GK2}]
  The length of the transient to reach the fixed point starting from
  the initial configuration $(n)$ for SPM is given by the following
  formula
  $$t_{seq} = \frac{1}{6}(p+1)p(p-1) + \frac{1}{2}k(2p+1-k) =
  \mathcal{O}(n^{3/2}) \enspace ,$$
  where $\para{p,k}$ is the decomposition of $n$ in its integer sum.
\end{theorem}

When parallel execution mode is used for SPM, things are a little bit
more complex and one can only give an upper and a lower bound for the
transient length.

\begin{theorem}[\cite{GK2}]
  In the parallel execution mode, the length of the transient to reach
  the fixed point starting from the initial configuration $(n)$ for SPM
  can be bounded as follows
$$
\mathcal{O}(n)=\frac{t_{seq}}{p-1}\leq t_{par}\leq t_{seq}=\mathcal{O}(n^{3/2})
\enspace,
$$
where $\para{p,k}$ is the decomposition of $n$ in its integer sum.
\end{theorem}

\medskip
Finally for \IPM, all the paths from $(n)$ to the fixed point do not
necessarily have the same length, but the longest chain in sequential
mode is known. Its exact expression can be found in~\cite{GMP1}.


\section{Generalization to arbitrary sandpiles} \label{sec:algo}
The results from the previous section are very complete, but they only
apply to the particular case of initial configurations of
length one.

A little more general study was started in~\cite{GK2}, where the
authors remarked that when starting from decreasing configurations
(for all $0 \leq i < l$, $a_i \geq a_{i+1}$) the lattice structure is
maintained with SPM. This result is extended in~\cite{P}, where the
author exhibits the set of fixed points but she does not associate to
each initial configuration the corresponding fixed point.

In this paper, we try to generalize these results to arbitrary initial
configurations giving a fast algorithm for computing fixed points.

\bigskip
\noindent \textbf{Description of the algorithm.} It consists in the
iteration of two major steps:
\begin{description}
\item[\texttt{Cut}:] the configuration is divided into intervals so that the
 formulas of the next step can be applied;

\item[\texttt{Compute}:] each interval is analyzed trying to figure out
  how it will be within a few iterations steps of the model.
\end{description}

Remark that it is necessary to iterate since some grains can pass from
one interval to another. Consider the case of two isolated columns of
grains $(m, 0, \ldots, 0, n)$. The first column will collapse.  Then,
depending on the model chosen and on the value of $m$, $n$ and of the
gap between the two columns, the grains of the first column will be
blocked by the second one or they will partially cover it.

\medskip

From now on, to simplify notations and proofs the results will be
given only for SPM. Their extension to other models such as \IPM is
straightforward, using the formulas given in~\cite{GMP1}. In fact our
results extend to all models satisfying two conditions: \emph{lattice
  structure} of the orbit graph, and \emph{reachability} detection
(one must be able to say whether a given configuration is reachable or
not, in a way similar to Lemma~\ref{lemm:cond1}). Hence we will
implicitly consider this class of sandpile models in the sequel of
the paper when talking about ``all'' models. Next section justifies
the restriction to these models.

\subsection{Cut}

The way we are going to split the configuration into intervals is very
simple: each interval has to contain a configuration which is
reachable by the model, and it is the longest satisfying this property.

At that point, the previous results need to be extended. Indeed, their
authors supposed that the grains could move as far as possible. In the
present case, the movement is limited to fixed intervals, grains are
prevented from going too far on the right. All the results remain
true, but have to be reformulated.

\begin{lemma}\label{lem.nocycle}
For all initial configurations $c$ for SPM, $\og{c}$ has no cycles.
\end{lemma}
\begin{proof}
  Given a configuration $c=(a_1, \ldots, a_l)$, consider the quantity
  $\varphi(c) = \sum_{i=1}^l \sum_{j=1}^{a_i} j$. If $c'=(a'_1,
  \ldots, a'_{l'})$ is obtained from $c$ by applying the SPM rule at
  column $i \leq l$, then $\varphi(c') = \varphi(c) - a_i + a'_{i+1} =
  \varphi(c) - a_i + a_{i+1} + 1$. Since the rule could be applied,
  $a_i \geq a_{i+1}+2$, which implies $\varphi(c') < \varphi(c)$; 
  in other words, the SPM system rule is irreversible.
  \qed
\end{proof}

Our algorithm will only work with models having this
\emph{irreversibility} property. \IPM has it, and any realistic model
should have it.

\begin{lemma}\label{lem.fixpunique}
For all initial configurations $c$ of length at most $l$ for SPM, 
$\og{c}^l$ has a unique fixed point $\Pi$. Moreover, $\Pi\in\og{c}$.
\end{lemma}
\begin{proof}
  Let $c$ be a configuration of length at most $l$, with $n$ grains.
  First of all remark that $\og{c}^l$ is a sub-graph of $\og{c}$ since
  for every configuration in $\og{c}^l$, the path from the root $(n)$
  to this configuration is also in $\og{c}$ (it just consists in
  applying system rules to columns of index less than $l$).
  
  Hence $\og{c}^l$ is finite, and by Lemma~\ref{lem.nocycle}, it has
  no cycles. Therefore every path starting from a configuration $c$
  reaches a fixed point. Assume that $\Pi_1$ and $\Pi_2$ are two
  distinct fixed points of $\og{c}^l$. They also belong to $\og{c}$,
  so they satisfy Lemma~\ref{lemm:cond1} \ie they are decreasing and
  contain at most one plateau (as they are fixed points of SPM, there
  are no cliffs). Their structure is represented on
  Figure~\ref{fig:struct}.  They are simply ``staircases'' with the
  addition of $k$ grains:
  $$
   \Pi_j = \left\{\begin{array}{l@{\hspace{2mm}}l}
      (p_j, p_j-1, \ldots, p_j-l+1) & \textrm{if $k_j=0$}\\
      (p_j, p_j-1, \ldots, p_j-\alpha_j, p_j-\alpha_j, \ldots, p_j-l+2) &
      \textrm{otherwise},
    \end{array}\right.
   $$
  with $\alpha_j = l - k_j -1$.
  \begin{figure}[!ht]
    \centering
    \input{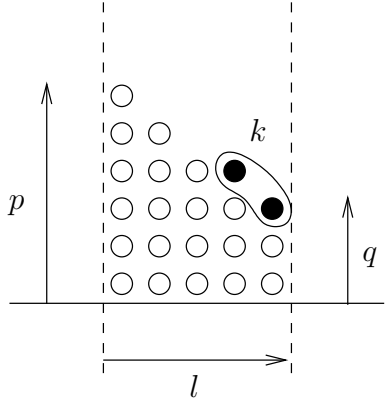}
    \caption{structure of a fixed point in $\og{c}^l$. The symbol $q$
    will be defined in Section~\protect\ref{sec.compute}.}
    \label{fig:struct}
  \end{figure}

  Counting the number of grains in $\Pi_1$ and $\Pi_2$ gives
  \begin{equation}\label{equa:count}
    \begin{array}{rcl}
      n &=& \displaystyle\sum_{i=1}^l (p_1+1-i) + k_1 =
      \displaystyle\sum_{i=1}^l (p_2+1-i) + k_2 \qquad(0 \leq k_1, k_2
      < l)\\[0.5ex]
      &=& \dfrac{1}{2}l(2p_1+1-l) + k_1 = \dfrac{1}{2}l(2p_2+1-l) +
      k_2\enspace.
    \end{array}
  \end{equation}
  This implies that $|p_2 - p_1| = \frac{1}{l}\cdot|k_2 - k_1| < 1,$
  and as $p_1$ and $p_2$ are integers we have $p_1=p_2$. It follows
  that $k_1=k_2$, and $\Pi_1=\Pi_2$.  \qed
\end{proof}

The irreversibility condition needed to apply the algorithm to a
particular model can be relaxed. In fact, what is needed for the
previous lemma to work is a lattice structure. This ensures
irreversibility and unicity of the fixed point.

\begin{proposition} \label{prop:sstreillis}
For all initial configurations $c$ of length at most $l$ for SPM, 
$\og{c}^l$ is a lattice and $\og{c}^l\subseteq\og{c}$.
\end{proposition}
\begin{proof}
  In the proof of Lemma~\ref{lem.fixpunique}, we saw that 
  $\og{c}^l$ is a sub-graph of $\og{c}$.  By
  Lemma~\ref{lem.nocycle} and~\ref{lem.fixpunique} we have the thesis.
  \qed
\end{proof}

\begin{lemma} \label{lemm:cond2}
  Consider the SPM model, a configuration $c$ of length at most $l$
  and let $n$ be its number of grains. Then, $c\in\og{(n)}^l$ if and
  only if it is decreasing and between any two plateaus there is at
  least a cliff.
\end{lemma}
\begin{proof}
  Consider a configuration $c\in\og{(n)}^l$, where $n$ is the number
  of grains of $c$. By Proposition~\ref{prop:sstreillis}, $c$ belongs
  also to $\og{(n)}$ and hence it satisfies the hypothesis of
  Lemma~\ref{lemm:cond1}.

  For the opposite implication, let $c=(a_1,\ldots,a_k)$ be a
  decreasing configuration of length at most $l$ in which any two
  plateaus have a cliff in between. Let $n$ be the number of grain of
  $c$. By Lemma~\ref{lemm:cond1}, $c$ belongs to $\og{(n)}$. If we
  consider the path going up to the root, all the configurations
  encountered are of length less than $l$ since the application of the
  system rule can only increase the length of a configuration. All the
  applications of the system rule take place in the interval $[1,
  l-1]$, hence each element of the path is also in $\og{(n)}^l$.
  \qed
\end{proof}

The previous lemma underlines the second condition that is needed for
the sandpile model used in the algorithm: it should be possible to
detect whether a configuration is \emph{reachable} by the model.
Moreover, for complexity issues (see Section~\ref{sec:compl}), it
should be as fast as possible (for SPM and \IPM, it is linear, as a
single scan suffices).

Using the previous lemma one can complete the characterization of
$\og{}^l$ that begun with Proposition~\ref{prop:sstreillis}.

\begin{proposition}\label{prop.graph-subgraph}
  For any integer $n$ it holds that
  $$\og{(n)}^l = \og{(n)}\big[\{c \in \N^l\}\big]\enspace,$$
  where for any graph $G=\para{V,E}$,  $G[V']$ is the sub-graph generated 
  by the set of vertices $V'\subseteq V$.
\end{proposition}

The previous proposition means that for SPM, $\og{(n)}^l$ is exactly
the sub-graph of $\og{(n)}$ restricted to the configurations of length
at most $l$, keeping all edges between these configurations. By
Proposition~\ref{prop:sstreillis}, this sub-graph is also a lattice.
This holds for any model having the lattice structure and the
reachability detection, this is the case for example for \IPM.

\medskip
The \texttt{cut} step is performed using Lemma~\ref{lemm:cond1},
Propositions~\ref{prop:sstreillis} and~\ref{prop.graph-subgraph}.  A
scan is performed in order to find the maximal intervals in which the
corresponding sub-configurations are in $\og{c}$. For example, for
SPM, a new interval starts whenever there are two consecutive columns
$i$ and $i+1$ such that $a_i<a_{i+1}$, or there is a second plateau
not separated by a cliff (see Listing~\ref{algo.cuts}).

\begin{lstlisting}[frame=tB,escapechar=!,caption={procedure for cutting
a configuration into intervals using SPM.},label=algo.cuts]
procedure cut (c[]) {   // c is the initial configuration 
   nbp = 0;  // number of plateaus
   I = !$\emptyset$!;    // set of right extremities of the intervals

   for (i=1; i<l; i++) {
      if (c[i+1] > c[i]) {              // increase
         I = I !$\cup\;\;\{$!i!$\}$!;
         nbp = 0;
      } else if (c[i] - c[i+1] >= 2) {  // cliff
         nbp = 0;
      } else if (c[i+1] == c[i]) {      // plateau
         nbp ++;
         if (nbp == 2) {
            I = I !$\cup\;\;\{$!i!$\}$!;
            nbp = 0;
         }
      }
   }
   return I;
}  
\end{lstlisting}

By Lemma~\ref{lemm:cond1}, the sub-configurations given by the intervals
are in $\og{c}$ and they are ``maximal''. Moreover, 
by Propositions~\ref{prop:sstreillis} and~\ref{prop.graph-subgraph}, we 
know that each sub-configuration is in $\og{(n_i)}^{l_i}$ and that it reaches 
the fixed point of $\og{(n_i)}^{l_i}$, where

\begin{itemize}
\item $c_i$ is the configuration corresponding to the  $i^{\textrm{\scriptsize
      th}}$ interval, $c_i = (a_k)_{k \in I_i}$\enspace;
\item $l_i = |I_i|$ is the length of $c_i$\enspace;
\item $n_i$ is the number of grains of $c_i$\enspace.
\end{itemize}
Remark that the quantities $l_i$ and $n_i$ can be computed inside the
procedure \texttt{cut}.

\begin{figure}[!ht]
  \centering
  \includegraphics{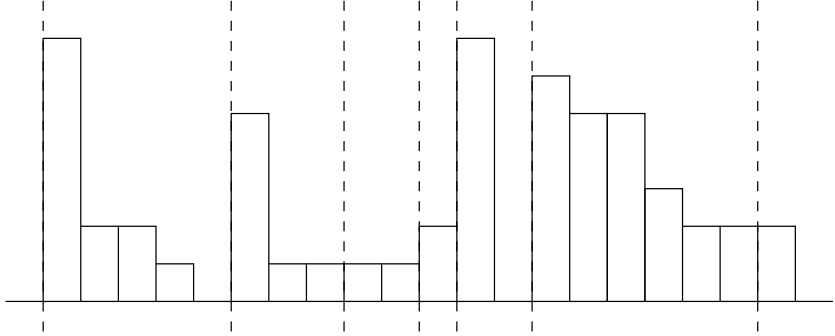}
  \caption{example of intervals after the \texttt{cut} step, for SPM.}
  \label{fig:decoup}
\end{figure}

The last interval -- reached when the scanning procedure arrives at
$a_l$ -- is a ``special case'': it is treated exactly like in the
usual model, supposing there is as much space as necessary.  An
example of ``cut'' for the SPM model is shown in
Figure~\ref{fig:decoup}.

\subsection{Compute}\label{sec.compute}

In this section we show formulas for finding the fixed point locally
to each interval. From now on and only for this section, we consider a
generic interval and hence the index $i$ will be omitted in the
notations. Again, only SPM will be considered to simplify the
formulas, but similar results hold for other models.

\medskip
The structure of the global configuration after the \texttt{cut} has
to be precisely determined: we shall find the fixed point of every
configuration in each interval.

In the proof of Lemma~\ref{lem.fixpunique}, we saw that the fixed
point $\Pi$ of an interval with SPM is a kind of ``staircase'' as
depicted in Figure~\ref{fig:struct}. We need the expression of $p$
(height of the leftmost column of the fixed point), $q$ (height of the
rightmost column) and $k$ (number of exceeding grains) in terms of $n$
and $l$.

Equation~\eqref{equa:count} implies that 
$$p = \left\lfloor \frac{n}{l} + \frac{l}{2} - \frac{1}{2}
\right\rfloor \enspace.$$
Then, $k=n-\frac{1}{2}l(2p+1-l)\enspace.$ For $q$,
the same calculation is done, counting the number of grains in the fixed point
$$\begin{array}{rcl}
  n &=& \displaystyle\sum_{i=1}^l (q+l-i) - k'\qquad(k' = l-k \mod l,
  \quad\textrm{hence } 0 \leq k' < l)
\end{array}
$$
one finds 
$q=\left\lceil\dfrac{n}{l}-\dfrac{l}{2}+\dfrac{1}{2}\right\rceil$.

\begin{remark} \label{rem:k}
  The quantity $k$ is not necessarily a real number of grains. Indeed,
  when the fixed point finishes by $(\ldots, 2, 1, 0, 0)$, we have
  $k=1$. This does not correspond to a grain, but compensates a
  negative grain introduced when computing $p$ (the last term of the
  sum equals $-1$).
  
  Moreover, the last interval has to be treated in another way, as it
  has infinite length. Choosing $l=\left\lceil \frac{\sqrt{8n+1}-1}
    {2} \right\rceil$, value obtained from the results in~\cite{GK2},
  would solve the problem. It corresponds to the length of the fixed
  point obtained from the configuration $(n)$.  With this new value,
  the calculation of $p$ and $q$ is correct.
\end{remark}

\subsection{Correctness}

In this section we will prove that the algorithm finishes and gives
the correct fixed point. All results are stated for all suitable
models. Proofs are given for the SPM case only, for simplicity sake.

The superscript $^f$ for the quantities $I_i,c_i,n_i$ mean that we use
their value they have at the end of the algorithm.

\begin{proposition} \label{prop:term}
  For any configuration, the algorithm finishes and returns a
  fixed point.
\end{proposition}
\begin{proof}
  Because of the irreversibility (Lemma~\ref{lem.nocycle}) and of the
  finite number of grains, the algorithm finishes.
  
  Suppose that it does not return a fixed point, \ie that it returns a
  configuration $c=(a_i)$ where there is an index $i$ such that a
  grain from $a_i$ can move to $a_{i+1}$. Then $a_i \geq a_{i+1}+2$,
  which implies that after the next \texttt{cut} step, $a_i$ and
  $a_{i+1}$ will belong to the same interval $I$ (by definition of the
  \texttt{cut}). As there is a possible evolution in $I$, the
  algorithm has to compute the new configuration where at least this
  grain moved.
  \qed
\end{proof}

We have to make sure that the fixed point characterized by the
previous proposition is the ``right'' fixed point.

The following lemma is quite straightforward for SPM. For other
models, the proof depends on the model itself, although it comes from
the irreversibility. For example for \IPM, there will be more cases to
consider but the idea remains the same.
\begin{lemma}\label{lemm:descendent}
  If a grain can move past a column $i$, then if any movement of grain
  passing this column is blocked, at least one of these movements will
  always remain possible.
\end{lemma}
\begin{proof}
  In the case of SPM the lemma can be reformulated as follows: a cliff
  at column $k$ remains until the system rule is applied at column
  $k$.

  Suppose that the SPM system rule can be applied to a configuration
  $c=(a_1, \ldots, a_l)$ at column $k$. It means that $a_k \geq
  a_{k+1} + 2$. If the rule applies at position $j \neq k$, and if
  we note $c'=(a'_1, \ldots, a'_{l'})$ the new configuration, there are
  four cases:
  \begin{itemize}
    \item if $j < k-1$, $a'_k = a_k$ and $a'_{k+1} = a_{k+1}$;
    \item if $j = k-1$, $a'_k = a_k+1$ and $a'_{k+1} = a_{k+1}$;
    \item if $j = k+1$, $a'_k = a_k$ and $a'_{k+1} = a_{k+1}-1$;
    \item if $j > k+1$, $a'_k = a_k$ and $a'_{k+1} = a_{k+1}$\enspace.
  \end{itemize}
  In all cases, $a'_k \geq a_k \geq a_{k+1}+2 \geq a'_{k+1}+2$. Hence
  the rule can still be applied at column $k$.
  \qed
\end{proof}

\begin{theorem}
  For any configuration $c$, $\og{c}$ is a lattice and its fixed point
  coincides with the fixed point found by the algorithm.
\end{theorem}
\begin{proof}
  Consider the intervals $I_i^f$ obtained after the application of our
  algorithm to $c$ . Split $c$ according to these intervals
  (remark that we do not use the intervals given by the first \texttt{cut}
  step, we only use the final ones) into sub-configurations 
  $c_k=(a_k)_{k\in I_i^f}$.
  
  No grain of any $c_i$ ever moves to another $c_j$. Suppose it is not
  the case; we are going to simulate the behavior of the model, but
  inhibiting the applications of the system rule which move grains
  from an interval to another, raising a contradiction.  Such a
  simulation ends when all the partial configurations $c_k$ reach a
  fixed point.  Let $I_i^f$ be an interval in which $c_i$ was once
  supposed to lose a grain. At this point of the simulation, the rule
  can still be applied in $I_{i}^f$ (Lemma~\ref{lemm:descendent}).
  Moreover, the fixed points of $c_{i}$ and $c_{i+1}$ correspond
  respectively to $c_{i}^f$ and $c_{i+1}^f$ computed by the algorithm,
  because they belong to two different orbit graphs (lattices) and
  they do not receive nor lose any grain (we recall that the rules
  moving grains from one interval to another have been inhibited).
  Therefore, the rule should also be applicable for our algorithm at
  the same column in $I_{i}^f$, which is not the case otherwise the
  algorithm would not have returned (Lemma~\ref{prop:term}).
  
  Therefore, the behavior of each $c_k$ is not influenced by the
  others, and hence the set of reachable configurations can be
  obtained by joining all the possible behaviors for every $I_i^f$.
  Thus $\og{c}$ is a lattice since it is the product of all the
  lattices $\og{(n_i^f)}^{l_i^f}$.

  The fixed point found by the algorithm is necessarily a fixed point
  of $\og{c}$. Hence it coincides with the fixed point of $\og{c}$
  since $\og{c}$ is a lattice.
  \qed
\end{proof}

\subsection{Complexity} \label{sec:compl}

Our algorithm is a loop divided into two parts: the \texttt{cut} step
and the \texttt{compute} step.  In the \texttt{cut} step, the initial
configuration is scanned and its complexity is $\mathcal{O}(l)$.
Moreover, we can assume that the \texttt{compute} step is done in
constant time for each interval, hence in $\mathcal{O}(l)$ for the
entire configuration (there are at most $l$ intervals, for a strictly
increasing configuration for example).

The number of iterations of the loop is a little harder to
evaluate. Intuitively, if there are many intervals, which means that
the configuration is mostly non decreasing, the fixed point will be
reached quite soon. If there are few intervals, then there will be few
iterations because all the calculi will be done in the
\texttt{compute} step. This is difficult to formalize in the general
case, all we can do for now is give a quite large upper bound.

\begin{proposition}
  The algorithm performs at most $\frac{1}{2} \cdot l \cdot
  (l+2f(n)-1)$ iterations, where $f(n)$ is the length of the fixed
  point reached by the configuration $(n)$ (for example, $f(n) =
  \lceil (\sqrt{8n+1}-1)/2 \rceil = \mathcal{O}(\sqrt{n})$ for SPM,
  see Remark~\ref{rem:k}).
\end{proposition}

\begin{proof}
  First, note that there are at most $l$ intervals. Consider the
  changes in the bounds of the intervals (and not in terms of grain
  content) between two iterations, \ie between two \texttt{cut} steps.
  When there is no change, the algorithm returns. Hence at least one
  of the intervals ``evolves'' at each iteration, except the last one.

  Consider the upper bound (highest index) $u_i$ of the interval $I_i$
  at the beginning of the program. Clearly, $u_i \geq i$, hence $u_i$
  can increase at most $l+f(n)-i-1$ times. It can increase one by one,
  until it reaches the end of the configuration or it disappears.
  $l+f(n)-1$ is the maximum length of a configuration of size $l$ with
  $n$ grains, it is obtained when nearly all grains are in the last
  column. Since at least one $u_i$ increases at each step (except the
  last step), there are at most
  $$\sum_{i=1}^{l} (l+f(n)-i-1)= \sum_{i=f(n)-1}^{l+f(n)-2} i =
  \frac{1}{2} \cdot l \cdot (l+2f(n)-3)$$
  changes of the intervals.

  Add $l-1$ iterations for the loss of intervals, plus one final
  iteration to detect that nothing happened, and the result follows.
  \qed
\end{proof}

Each iteration has a complexity of $l$ (provided the \texttt{cut} step
is linear and the \texttt{compute} step is constant, which is the case
for SPM and \IPM), hence the global complexity in the general case is
in $\mathcal{O}(l^2 \cdot (l+2f(n)))$. This is not so interesting,
because it does not apply to any model in particular. Therefore it is
not possible to give a sense to this result, comparing it to previous
results.

But it will be refined in the next section in the special case of SPM,
and then it will be possible to compare it to the classical simulation.


\section{Improvement for SPM} \label{sec:spm}

The algorithm described above works for the main existing sandpile
models, provided the model has a lattice structure in which every
element can be easily characterized. This ensures unicity of the fixed
point, and the characterization enables to cut the configuration, to
compute the fixed point faster.

The simplicity of SPM allows a few optimizations.

\subsection{Merge}

\begin{figure}[!ht]
  \centering
  \includegraphics{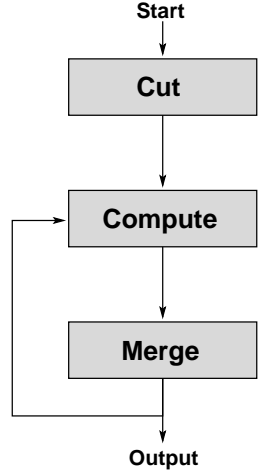}
  \caption{structure of the algorithm refined for SPM}
  \label{fig:algoSPM}
\end{figure}

The main optimization which can be achieved with SPM is the removal of
the iteration over the \texttt{cut} step. The scan of the
configuration can be replaced by a scan of the intervals in order to
\emph{merge} successive intervals when possible. The new structure of the
algorithm is represented in Figure~\ref{fig:algoSPM}.
This operation does not affect the theoretical complexity, as the
number of intervals can be as high as the number of columns, but in
practice the gain is very important (very few intervals are actually
of size 1, and they tend to disappear upon iterations).

\medskip
At each iteration of the \texttt{merge} step, the algorithm tries
to combine intervals by considering the difference of height at their
borders. If the merge succeeds, then new values are computed for
the new interval.

When looking at the border between two intervals $I_i$ and $I_{i+1}$,
there are two possibilities.
\begin{itemize}
\item Either $q_i \leq p_{i+1}+1$, in which case nothing can happen at
  the border. There are no cliffs inside the intervals by construction
  of the fixed point, hence nothing has to be done.
  
\item Or $q_i \geq p_{i+1}+2$, which means that the system rule can be
  applied at the border. To obtain a new fixed point, the two
  intervals will be merged into $I'_{\alpha(i)} = I_i \cup I_{i+1}$,
  where $\alpha$ is some suitable renumbering function. Remark that
  from one iteration to another, some intervals disappear (merge) and
  some remain, introducing lag in the indices.
  
  By Lemma~\ref{lemm:cond2}, in this new interval the configuration
  belongs to $\og{(n_i+n_{i+1})}^{l_i+l_{i+1}}$. Indeed, there is at
  most one plateau in $I_i$, at most another one in $I_{i+1}$, and
  they are separated by the cliff at the border. Therefore, the 
  previous formulas still hold in the new interval $I_{\alpha(i)}$ 
  with
  $$\begin{array}{rcl}
    n'_{\alpha(i)} &=& n_i + n_{i+1}\\
    l'_{\alpha(i)} &=& l_i + l_{i+1}\\
    c'_{\alpha(i)} &=& (c_i, c_{i+1})\enspace.\\
  \end{array}$$
  Due to the fact that we are in the lattice $\og{(n'_{\alpha(i)})}
  ^{l'_{\alpha(i)}}$, we know that the computed fixed point will be
  correct. Therefore this operation can efficiently merge two
  intervals.
\end{itemize}

Once two intervals have merged, the interval list has to be updated
with the new interval which replaces the two old ones. That way, the
next iteration will act on the new list. This allows a newly created
interval to merge at the next step, letting grains move as far as
possible.

\subsection{Transient length}

For SPM, the algorithm can also compute the transient length. It is
not possible in general because all paths from the root to the fixed
point may not have the same length (\IPM for example), but it is the
case with SPM. So the number of steps simulated by the algorithm will
correspond to the transient length, independently from the choice of
the intervals.

To compute its value inside an interval, we associate to every sand
grain a number corresponding to its movement, as in~\cite{GK2}. All
these elementary values are added, supposing the configuration in
Figure~\ref{fig:struct} has been reached. Then we have to subtract the
``movement'' weight $t^0$ of the initial configuration. One finds
\begin{equation}\label{eq.comp.trans}
  t = \sum_{i=0}^{l-1} i(p-i) + \sum_{i=l-k}^{l-1}i - t^0
    = \frac{l(l-1)(3p-2l+1)}{6} + \frac{k(2l-k-1)}{2} - t^0\enspace,
\end{equation}
where $t^0=\sum_{i=0}^{l-1}i\cdot a_{i+j}$ can be computed during the
\texttt{cut} step; $j$ is the index of the leftmost column of
the interval under consideration.

\begin{remark}
  Again, the quantity $k$ is not really a number of grains. The
  rightmost sum of Equation~\eqref{eq.comp.trans} can be too big, but
  this will be compensated by a smaller value on the left. The extra
  grain added on the right is counted as a negative grain on the left.
\end{remark}

One has to add all these values, over each interval at each iteration,
to obtain the total transient length. When intervals do not merge,
$t$ does not change. Otherwise, we are going to count the number of
steps simulated, and subtract everything that had already been done
during last iteration.
$$\begin{array}{@{}rcl@{}}
t'_{\alpha(i)} &=& \dfrac{1}{2} l'_{\alpha(i)}
\left(l'_{\alpha(i)}-1\right)
\left(3p'_{\alpha(i)}-2l'_{\alpha(i)}+1\right) + \dfrac{1}{2}
k'_{\alpha(i)} \left( 2l'_{\alpha(i)} - k'_{\alpha(i)} - 1
\right)\\[1ex]
 &-& t_i - t_{i+1} - n_{i+1} \cdot l_i
\end{array}
$$
The last term is the number of steps spent to move $n_{i+1}$ grains
from interval $I_i$ to $I_{i+1}$.

\subsection{Summary}

Table~\ref{tab:recap} sums up all the calculations that have to be
computed at each iteration, in the SPM example. The notation
$\beta_i^j$ represents the value of the variable $\beta$ in the
$i^{\textrm{\scriptsize th}}$ interval at the $j^{\textrm{\scriptsize
    th}}$ iteration.

\begin{table}[!ht]
  $$\begin{array}{@{}c||c|c@{}}
    & \textrm{\textbf{A}} & \textrm{\textbf{B}}\\ \hline \hline
    I_{\alpha_j(i)}^{j+1} & I_{i}^{j} & I_{i}^{j} \cup I_{i+1}^{j}\\
    \hline
    n_{\alpha_j(i)}^{j+1} & n_i^j & n_i^j + n_{i+1}^j\\ \hline
    l_{\alpha_j(i)}^{j+1} & l_i^j &
    \left\{\begin{array}{l@{\hspace{0.5cm}}l}
        
        \left\lceil\dfrac{1}{2}\left(\sqrt{8n_{\alpha_j(i)}^{j+1}+1}-1\right)
        \right\rceil &
        \textrm{if $I_{i+1}^j$ is the last interval}\\[2ex]
        l_i^j + l_{i+1}^j & \textrm{otherwise}
      \end{array}\right.\\ \hline
    p_{\alpha_j(i)}^{j+1} & p_i^j & \left\lfloor
      \dfrac{n_{\alpha_j(i)}^{j+1}}{l_{\alpha_j(i)}^{j+1}} +
      \dfrac{l_{\alpha_j(i)}^{j+1}}{2} - \dfrac{1}{2} \right\rfloor\\ \hline
    q_{\alpha_j(i)}^{j+1} & q_i^j & \left\lceil
      \dfrac{n_{\alpha_j(i)}^{j+1}}{l_{\alpha_j(i)}^{j+1}} -
      \dfrac{l_{\alpha_j(i)}^{j+1}}{2} + \dfrac{1}{2} \right\rceil\\ \hline
    k_{\alpha_j(i)}^{j+1} & k_i^j & n_{\alpha_j(i)}^{j+1} -
    \dfrac{1}{2}l_{\alpha_j(i)}^{j+1} \left(2p_{\alpha_j(i)}^{j+1} + 1 -
      l_{\alpha_j(i)}^{j+1}\right)\\ \hline
    t_{\alpha_j(i)}^{j+1} & t_i^j & \begin{array}{l}
      \dfrac{1}{2} l^{j+1}_{\alpha_j(i)}
      \left(l^{j+1}_{\alpha_j(i)}-1\right)
      \left(3p^{j+1}_{\alpha_j(i)}-2l^{j+1}_{\alpha_j(i)}+1\right)\\[1ex]
      \qquad {} + \dfrac{1}{2} k^{j+1}_{\alpha_j(i)}
      \left(2l^{j+1}_{\alpha_j(i)} - k^{j+1}_{\alpha_j(i)} - 1 \right)
      - t_i^j - t_{i+1}^j - n_{i+1}^j \cdot l_i^j
    \end{array}
  \end{array}$$
\caption{summary of the operations performed by the algorithm modified
  for SPM, during one iteration. Column \textbf{A}
  contains the values obtained when $q_i^j \leq p_{i+1}^j+1$, and
  \textbf{B} the values for $q_i^j \geq p_{i+1}^j+2$.}
\label{tab:recap}
\end{table}

At the end of the execution, these values allow to determine
exactly the shape of the fixed point ($p$, $q$ and $k$ in each
interval), and the total transient length.

\subsection{Complexity (for SPM)}

As seen above, the theoretical complexity is not improved with the
modified algorithm, but we can refine the computation made in
Section~\ref{sec:compl}.

The \texttt{compute} and \texttt{merge} steps both consist in a scan
of the intervals, hence they are in $\mathcal{O}(l)$. The number of
iterations is clearly bounded by $\mathcal{O}(l)$, because there can
not be more merging than the number of intervals. Moreover, the
following proposition shows that it is also bounded by the number of
grains. This last constraint becomes interesting when grains are
scattered along the configuration, \ie when $l \gg n$.

\begin{proposition}\label{prop.compupbound}
  Consider a configuration with $n$ grains. The algorithm finds the
  fixed point performing at most $\mathcal{O}(n)$ \texttt{merge}
  steps.
\end{proposition}
\begin{proof}
  It suffices to prove that if there are $m$ mergings, then there are
  at least $m$ grains in the configuration. Indeed, in order for $I_i$
  and $I_{i+1}$ to merge we shall have $q_i \geq p_{i+1}+2 \geq 2$.
  This means that $I_i$ has at least two grains in its rightmost
  column, mark the lowest one. After the merging, the marked grain
  does not move because it is at height $1$; moreover, it is not
  anymore at the border of two intervals. Therefore, any successive
  merging will mark a different grain, which has not been marked yet.
  Thus, there are no more than $n$ mergings.
  \qed
\end{proof}

The following example shows that the bound given in
Proposition~\ref{prop.compupbound} can be reached.

\begin{figure}[!ht]
  \centering
  \includegraphics[width=\textwidth]{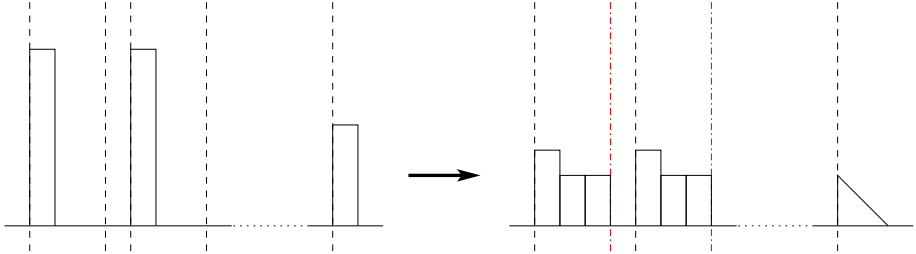} 
  \caption{example of configuration reaching the $\mathcal{O}(n)$ complexity}
  \label{fig:ctreex}
\end{figure}

\begin{example}
  Consider a configuration $c$ of length $l$ defined as follows
  $$
  \forall i\in\set{1,\ldots,l},\; c_i =
  \left\{\begin{array}{l@{\hspace{1cm}}l}
      7 & \textrm{if $i = 4j\ , \quad 0 \leq j < 4\lfloor n/7 \rfloor$}\\
      n \mod 7 & \textrm{if $i = 4\lfloor n/7 \rfloor$}\\
      0 & \textrm{otherwise.}
    \end{array}\right.
  $$
  In other words, $c = (7, 0, 0, 0, 7, 0, 0, 0, \ldots, 7, 0, 0, 0,
  n \mod 7)$ (figure~\ref{fig:ctreex}). The \texttt{cut} step produces
  $2\lceil n/7 \rceil - 1$ intervals of type $(7, 0, 0)$ and $(0)$.
  After the \texttt{compute} step, the configuration is $c' = (3, 2,
  2, 0, 3, 2, 2, 0, \ldots, 3, 2, 2, 0, x_1, x_2, x_3)$. One can
  easily see that there will be $\lceil n/7 \rceil- 1$ mergings of
  intervals of type $(3, 2, 2)$ and $(0)$ (in red, dashed dotted
  lines), in order to obtain the fixed point $(3, 2, 1, 1,$ $3, 2, 1, 1,
  \ldots, 3, 2, 1,1, x_1, x_2, x_3)$.  \qed
\end{example}

Our algorithm computes the fixed point of any initial configuration in
time $\mathcal{O}(l\cdot\min(n,l))$. There is a gain of at least
$\sqrt{n}$ compared to the naive simulation, whose complexity is
$\mathcal{O}(l\cdot n^{3/2})$ ($n^{3/2}$ is the number of avalanches,
while $l$ corresponds to the scan of the configuration in order to
find a cliff).


\section{Conclusions} \label{sec:concl}

In this paper we proposed an algorithm for finding the fixed points of
a large class of sandpile models starting from arbitrary finite
configurations.

Time complexity of the algorithm depends on the structure of the
initial configuration and, of course, on the model chosen. It might be
interesting to build hierarchies of models classified according to
time complexity of the (specialized version of the) algorithm, and
study the algebraic property of the induced orbit graphs.

Another research direction consists in the study of models which allow
grains to move in more than one direction and in parallel execution
mode. Remark in fact that grains in a sandpile are essentially subject
to two different type of forces: a vertical one due to gravity, and a
horizontal one due to wind. Realistic models should consider both
forces, and apply them in parallel to every grain. Our algorithm would
be adapted to such models, and would allow much faster computation of
this kind of natural phenomenon. Our research currently consists in
finding such a model, with the needed lattice structure.

\nocite*
\bibliographystyle{plain}

\end{document}